\documentclass[conference]{IEEEtran}
\IEEEoverridecommandlockouts

\usepackage[T1]{fontenc}
\usepackage{amssymb}
\usepackage{newtxtext,newtxmath} 

\usepackage[bookmarks=false]{hyperref}
\usepackage{graphicx,epsfig,amsfonts,amsmath,amssymb,mathrsfs,color,enumerate,empheq,mathdots,empheq}
\usepackage[noadjust]{cite}
\usepackage{booktabs}
\usepackage{placeins}
\usepackage{comment, balance}
\usepackage{stfloats}
\usepackage{floatrow}
\newfloatcommand{capbtabbox}{table}[][\FBwidth]
\usepackage{tikz}
\usepackage{placeins}
\usepackage[linesnumbered,vlined,ruled]{algorithm2e}
\usepackage{floatrow}
\newcommand{\subparagraph}{}
\usepackage{titlesec}
\usepackage[justification=centering]{caption}
\usepackage[usenames,dvipsnames]{pstricks}
\usepackage{hyperref}
\usepackage{dsfont}
\usepackage{multirow}
\usepackage{multicol}
\usepackage{subfig}
\usepackage{booktabs}
\usepackage{makecell}
\usepackage{siunitx}

\usepackage{floatrow}
\usepackage{tabularx} 
\usepackage{booktabs} 
\usepackage{ltablex} 
\usepackage{lipsum} 
\usepackage[belowskip=-30pt,aboveskip=0pt]{caption}

\def\BibTeX{{\rm B\kern-.05em{\sc i\kern-.025em b}\kern-.08em
    T\kern-.1667em\lower.7ex\hbox{E}\kern-.125emX}}

\DeclareSymbolFont{grb}{OML}{cmm}{b}{it}
\DeclareMathSymbol{\zetab}{\mathord}{grb}{"10}
\DeclareMathSymbol{\etab}{\mathord}{grb}{"11}
\DeclareMathSymbol{\thetab}{\mathord}{grb}{"12}
\DeclareMathSymbol{\kappab}{\mathord}{grb}{"14}
\DeclareMathSymbol{\lambdab}{\mathord}{grb}{"15}
\DeclareMathSymbol{\mub}{\mathord}{grb}{"16}
\DeclareMathSymbol{\nub}{\mathord}{grb}{"17}
\DeclareMathSymbol{\rhob}{\mathord}{grb}{"1A}
\DeclareMathSymbol{\sigmab}{\mathord}{grb}{"1B}
\DeclareMathSymbol{\taub}{\mathord}{grb}{"1C}
\DeclareMathSymbol{\phib}{\mathord}{grb}{"1E}
\DeclareMathSymbol{\psib}{\mathord}{grb}{"20}
\DeclareMathSymbol{\omegab}{\mathord}{grb}{"21}
\DeclareMathSymbol{\epsilonb}{\mathord}{grb}{"22}
\DeclareMathSymbol{\varphib}{\mathord}{grb}{"27}

\DeclareFontFamily{U}{mathx}{\hyphenchar\font45}
\DeclareFontShape{U}{mathx}{m}{n}{
      <5> <6> <7> <8> <9> <10>
      <10.95> <12> <14.4> <17.28> <20.74> <24.88>
      mathx10
      }{}
\DeclareSymbolFont{mathx}{U}{mathx}{m}{n}
\DeclareFontSubstitution{U}{mathx}{m}{n}
\DeclareMathAccent{\widecheck}{0}{mathx}{"71}

\newfont{\Bd}{msbm10 at 12 truept}
\newfont{\Sc}{eusm10 at 12 truept}

\def\diag{\mathop{\rm diag}\nolimits}

\def\0{{\bf 0}}
\def\1{{\bf 1}}

\def\0{{\bf 0}}

\def\bTheta{ {\boldsymbol \Theta} }

\DeclareCaptionJustification{justified}{\justified}

\makeatletter
\def\ifundefined{\@ifundefined}

\makeatother
\begin{document}

      \title{ 
      Extending Cislunar Communication Network Reach Using Reconfigurable Intelligent Surfaces}

    \author{ Aamer Mohamed Huroon\textsuperscript{‡*},
    Baris Donmez\textsuperscript{§},
     Gunes Karabulut Kurt\textsuperscript{§},
   and Li-Chun Wang\textsuperscript{‡}
   \\
\IEEEauthorblockA{
       \textit{  \textsuperscript{‡} Department of Electrical and Computer Engineering, National Yang Ming Chiao Tung University }\\
       \textit{\textsuperscript{{*}} 
           Department of Electrical and Electronic Engineering, University of Nyala } \\
       \textit{\textsuperscript{§} 
           Poly‐Grames Research Center, Dept. of Electrical Engineering, Polytechnique Montréal, Montréal, Canada}\\
       E-mail: \textsuperscript{‡}\{aamer,  wang\}@nycu.edu.tw} \textsuperscript{§}\{baris.donmez, gunes.kurt\}@polymtl.ca }
    \maketitle

     \begin{abstract}
This study introduces a novel approach to enhance communication networks in the Cislunar space by leveraging Reconfigurable Intelligent Surfaces (RIS). Using the ability of RIS to dynamically control electromagnetic waves, this paper tackles the challenges of signal attenuation, directivity, and divergence in Cislunar missions, primarily caused by immense distances and that Earth-based station transmitters do not always face the Moon. A new optimization problem is formulated, whose objective is to maximize the received signal-to-noise ratio (SNR) for Earth-to-Moon communications. We derive a closed-form solution to the problem of determining the optimal RIS phase shift configuration based on the effective area of the RIS.  Through extensive simulations, this paper demonstrates how optimal adjustments in RIS phase shifts can significantly enhance signal integrity, hinting at the substantial potential of RIS technology to revolutionize long-distance Cislunar communication.

    \end{abstract}


    \section{Introduction}\label{sec_into}
 In the quest to push the boundaries of human exploration, lunar missions stand as a testament to our technological ambition and prowess. These ventures into the vastness of space are not without significant challenges, chief among them the imperative of establishing robust communication links. These systems serve as the lifeline between the distant reaches of our solar system and Earth, facilitating the transmission of critical data and commands essential for the success and safety of such missions \cite{9982444}.
 
 The importance of communication in these endeavors cannot be overstated, particularly as we confront the need to maximize the signal-to-noise ratio (SNR). This shift in focus arises not only from technical constraints but also from a commitment to enhancing the quality of data transmission while simultaneously mitigating interference with radio astronomy a field that relies on the detection of faint celestial signals, which can be easily overshadowed by transmissions from lunar missions. The execution of lunar missions requires navigating complex dynamics, particularly the need to balance effective communication over maximum distances. Achieving this balance is crucial for mission success and can be accomplished by maximizing the SNR. Thus, striking a balance between communication effectiveness and the transmit power level is vital, underscoring the complex dynamics at play in the execution of lunar missions \cite{9982444, xu2019effects,  giordani2020non}. Hundreds of million Euros financial loss per year is estimated for the ground-based astronomical research facilities in which many researchers struggle to make observations due to the satellite constellations covering the Earth \cite{gallozzi2020concerns}. The radio-astronomy bands overlap with L, S, C, X, Ku, Ka, and V bands which are also used by low earth orbit (LEO) satellites for communication. Although this Radiofrequency Interference (RFI) is inevitable, the mitigation methods can decrease the RFI to tens of dB \cite{gallozzi2020concerns}. Therefore, the selection of the transmit power is crucial and can prevent the astronomers to make any observations within these spectrums.

Addressing these intricate challenges, this paper introduces an innovative approach through the deployment of Reconfigurable Intelligent Surfaces (RIS) \cite{toka2024ris,  puumala2023moving, 10333560, cetin2023secure, huroon2022generalized, donmez2024continuous, koktas2022communications}. By using RIS modules on a Geostationary Orbit (GEO) satellite, which are capable of reflecting signals from Earth to the Moon, as shown in Fig.~\ref{fig:system_dsc23}, we can demonstrate that GEO-mounted RISs.  This technology signifies a transformative advancement in Cislunar communication, addressing traditional challenges like signal attenuation and directivity issues that have impeded conventional frameworks. By leveraging RIS, it enables continuous and optimized phase rotation of electromagnetic waves, representing a paradigm shift in how we manage and improve communication signals in space.

This capability allows for precise control over the scattering, absorption, reflection, and diffraction of waves, thereby enhancing signal directivity, power, and network coverage. Our exploration into RIS and its applications in non-terrestrial networks, satellite broadcasting, and potential future high-altitude platform station (HAPS) networks illustrates the technology's versatility and its significant implications for global communication systems \cite{ koktas2023unleashing, tekbiyik2022reconfigurable, huroon2025multi,  9779261, huroon2024uav, bjornson2022reconfigurable,  cao2021reconfigurable}.

This work marks the first attempt at the integration of RIS \cite{7500635, koktas2022communications, koktas2023unleashing, 9779261} into Cislunar communication networks, revealing their critical role in improving signal quality through enhanced SNR, mitigating signal loss, and refining directivity. 
The deployment of RIS-assisted Cislunar communication faces several critical challenges. First, the high phase sensitivity of RIS and the need for precise alignment under dynamic and constantly varying lunar orbits make it difficult to maintain an optimal SNR. Second, hardware limitations combined with the complexities of real-time adaptive control further restrict the feasibility of practical implementation in such harsh and resource-constrained Cislunar environments.

The contributions of this paper are summarized as follows:
\begin{itemize}

\item The lunar environment is established realistically by utilizing the high-precision orbit propagator (HPOP), which considers third-body gravitational forces (e.g., Sun), solar radiation pressure, and central body radiation pressures (i.e., albedo, thermal) in Systems Tool Kit (STK)~\cite{donmez2025hybridfsorflunar}.
 A full Moon revolution around the Earth is considered in this work, which is crucial, because the results of a work with a limited simulation duration cannot be used straightforwardly for further projections, as the movement of the satellites change due to the aforementioned external forces, especially for simulations with longer time intervals and without the stationkeeping~\cite{donmez2025multiorbitercontinuouslunarbeaming}.    

    \item A new system model for Cislunar communications where ground-based Earth stations communicate with the Moon is proposed. We advocate using RIS-equipped GEO satellites to ensure reliable communication, considering that the position of the ground base station and beam divergence may result in the absence of a direct line-of-sight path.
    To take advantage of the proposed RIS-mounted GEO satellite for Cislunar communication assistance, we formulate a constrained optimization problem which seeks the optimal phase rotation that results in maximal SNR. Closed-form solutions are then obtained for the problem considered.

        
        \item Through extensive simulations, we gather substantial evidence demonstrating the effectiveness of the proposed RIS-mounted GEO satellite enhances the SNR by 3.2 dB through the use of a  RIS with 100 reflecting elements for Cislunar communication. Our results hint at the great potential of RIS in Cislunar communications, marking one of the first attempts at employing RIS in space communications.
       
\end{itemize}

 The structure of this paper is outlined as follows: Section~\ref{sec_bac} introduces the system model. Section~\ref{sec_meth} details the problem formulation. Section~\ref{sec_sim} discusses the simulation results. The paper is concluded in Section~\ref{sec_con}.
\begin{table}[b]
\caption{Notations used in the analysis}
\centering
\footnotesize
\begin{tabular}{cl}
\hline
Notation & Description \\
\hline
\( P_t \) & Power of the transmitted signal by the spacecraft. \\
\( P_r \) & Power of the received signal on the source. \\
\( G_t \) & Gain of the transmitting antenna. \\
\( G_r \) & Gain of the receiving antenna. \\
\( L_{fs} \) & Free-space path loss. \\
\( L_{ris} \) & Additional path loss due to RIS. \\
\( D \) & Distance between the source and the destination. \\
\( \psi\) & reflection angle.\\
\( N \) & Noise level at the receiver. \\
\( SNR \) & Signal-to-noise ratio at the receiver. \\
\( A_{eff} \) & Effective area. \\
\( d  \) & The distances from the source to the RIS. \\
\( d' \) & The distances from RIS to the destination. \\
\( \Theta_{ris} \) & RIS phase Configuration parameters. \\
\hline
\end{tabular}
\end{table}
    \section{System Model }\label{sec_bac}
    
\begin{figure}[!t]
    \centering
             \includegraphics[width=1.0\linewidth]{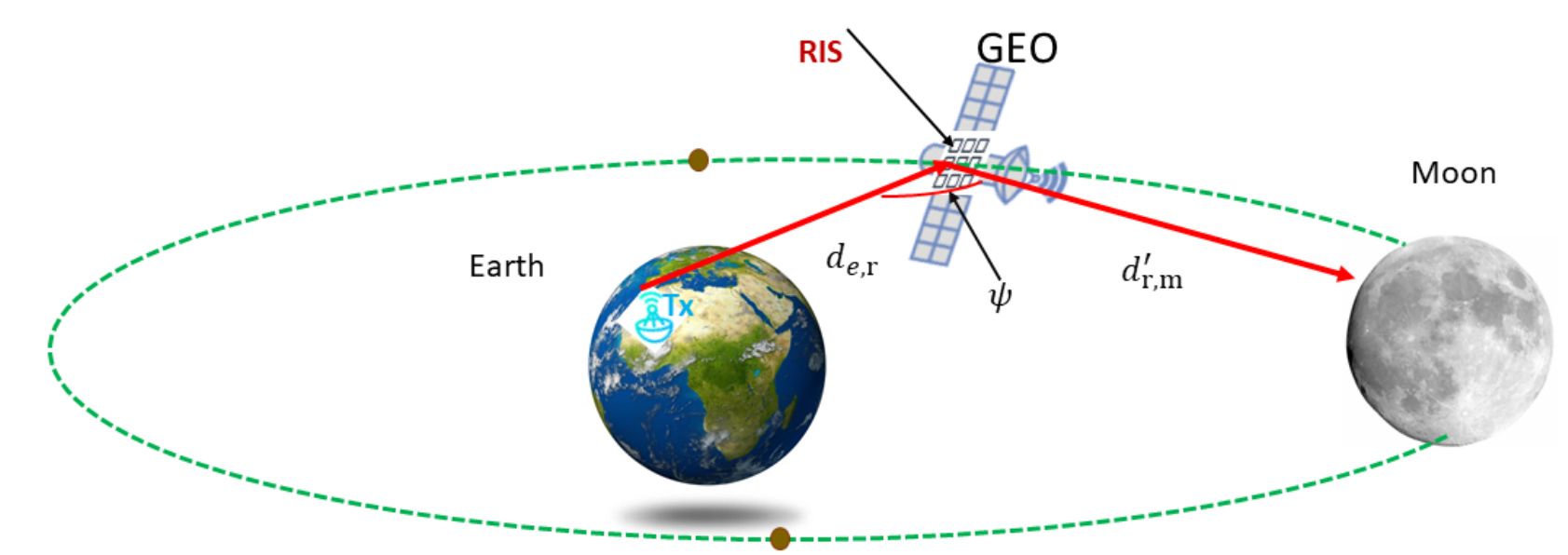}
                    \caption{ System Model for a Cislunar Communication Network.}
                    \label{fig:system_dsc23}
    
    \end{figure}

\begin{figure}[!b]
\centering
\subfloat[]{
	\label{subfig:STK1a}
	\includegraphics[clip, scale=0.28]{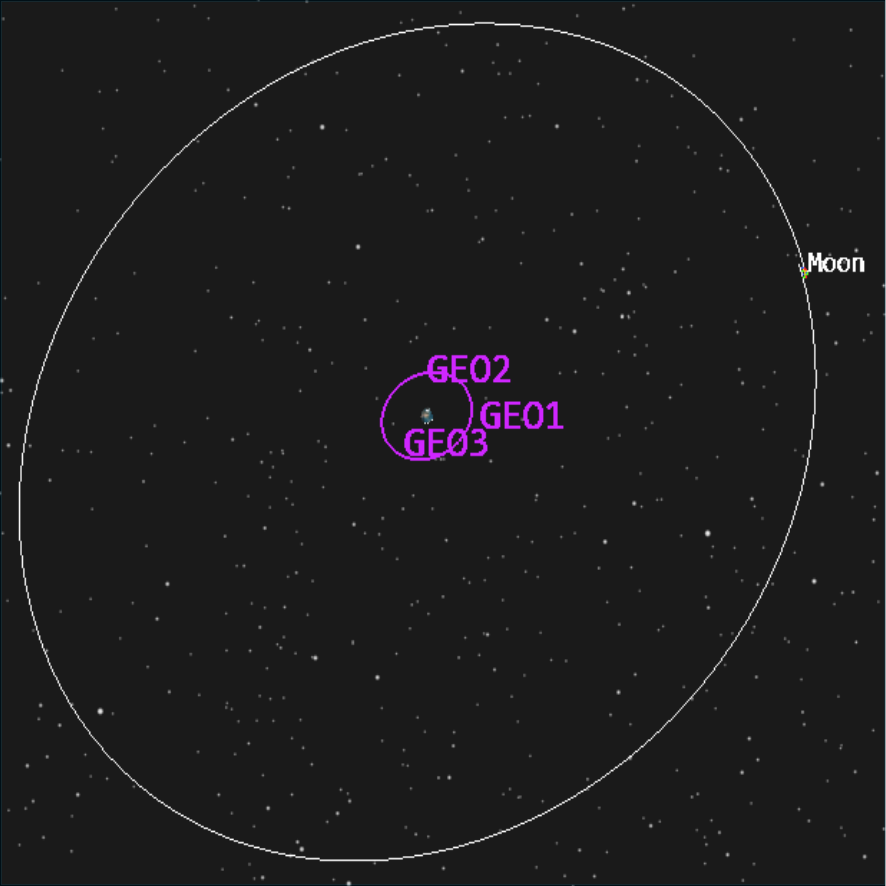}
	 }
\subfloat[]{
	\label{subfig:STK1b}
	\includegraphics[clip, scale=0.28]{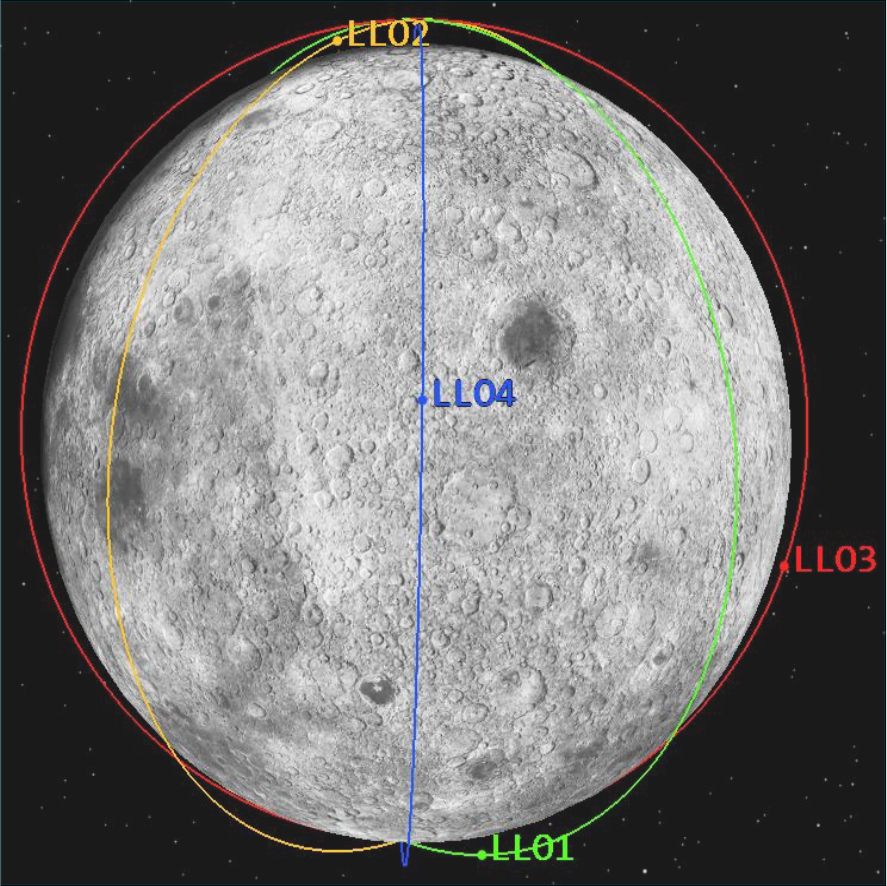}
	 }
\caption{Time-varying Cislunar environment in STK (a) GEO satellites rotating along with the Earth, (b) LLO satellites moving around the Moon.}
\label{fig:STK1}
\end{figure}

  In this section, we present the system model for RIS-assisted Cislunar communication, specifically designed to enhance the communication capabilities towards the Moon and beyond, as depicted in Fig.~\ref{fig:system_dsc23}. The model encompasses a communication source aiming to transmit signals to a distant destination on the Moon with the aid of a RIS-equipped GEO satellite. The model delineates the communication path into two segments: from the source to the GEO satellite over a distance $d$, and from the GEO satellite to the destination over a distance $d^\prime$. To optimize the communication range, we select the shortest link among multiple available links between the GEO satellites and LLO satellites, which are illustrated in 
  Fig.~\ref{fig:STK1}.

    The RIS technology utilized in this model is characterized by a passive surface with $M$ elements engineered to reflect signals directly toward the Moon. This approach leverages the inherent capability of RIS to manipulate signal paths effectively, thus enhancing the communication link's overall performance \footnote{There are many types of RIS architectures, including active, passive, and beyond diagonal RIS (BD-RIS). In this paper, as the first attempt, we focus on the passive RIS, the most practical one among them.}.

As demonstrated in Fig. \ref{fig:STK1}, three equidistant GEO satellites rotate together with the Earth, and four LLO satellites move around the Moon. When there is only one or two LLO satellites considered, they are positioned at the lunar far side simultaneously at some instants so a link outage occurs between the GEO and the relaying LLO satellites. To overcome the link outage, a single orbiter moving around each of the four LLOs is taken into account. The simulation results verify that at least a single link is available between the GEOs and LLOs in each time step during 27.3 days. Then, we applied a link selection algorithm \cite{donmez2025multiorbitercontinuouslunarbeaming} for the consideration of the shortest link.     

The Keplerian orbital parameters for both GEO and LLO satellites in our simulations over 27.3 days are presented in Table \ref{tab:Kepler}. The altitude values of the LLO satellites vary around 100 km due to the application of the time-varying external forces over the satellites.

    \begin{figure}[!b]
    \centering
             \includegraphics[width=0.9\linewidth]{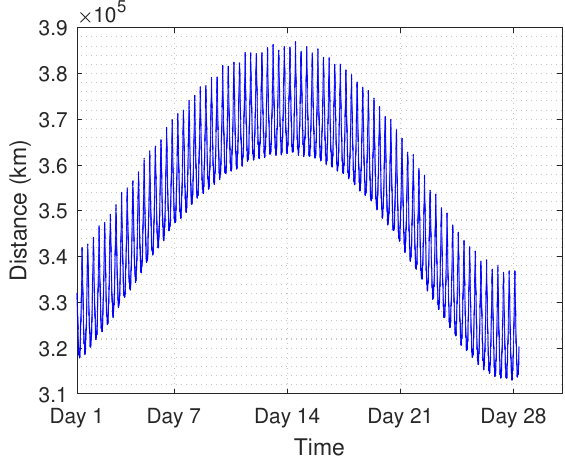}
                    \caption{Time-varying distance between the closest GEO satellite and LLO satellite.} 
                    \label{fig:system13}
    \end{figure}

\begin{table}[b]
  \sisetup{group-minimum-digits = 4}
  \centering
  \caption{Simulation Parameters in STK}
  \label{tab:Kepler}
  \resizebox{\columnwidth}{!}{%
  \begin{tabular}{lllS[table-format=5]ll} 
    \toprule
    \toprule
    \textbf{GEO satellites} \\
     Semi-major axis (Altitude) & 42,378.1 km (36,000 km) \\
     Eccentricity & $\approx0$   \\
     Inclination & $23.44^{\circ}$   \\
     Argument of perigee & $0^{\circ}$  \\
     Right ascension of the ascending \\ node (RAAN) & $0^{\circ}$ \\
     True anomaly & $0^{\circ}$, $120^{\circ}$, $240^{\circ}$  \\
     Satellite mass & 5000 kg  \\
    \midrule 
     \textbf{LLO satellites} \\
     Semi-major axis (Altitude) & 1837.4 km (100 km) \\
     Eccentricity & $\approx0$   \\
     Inclination & $90^{\circ}$   \\
     Argument of perigee & $0^{\circ}$  \\
     Longitude of the ascending node &  $0^{\circ}$, $90^{\circ}$, $225^{\circ}$, $315^{\circ}$ \\
     True anomaly & $0^{\circ}$, $180^{\circ}$, $90^{\circ}$, $270^{\circ}$  \\
     Satellite mass & 1000 kg  \\
     \midrule
     Total simulation time & 27.3 days \cite{donmez2025multiorbitercontinuouslunarbeaming}   \\
     Sampling time & 1 minute   \\
    \bottomrule
    \bottomrule
\end{tabular}
\vspace{-0.2 cm}
}
\end{table}

\subsection{ Signal Model}

    In RIS-assisted Cislunar communication, a ground base station on Earth initiates communication with the Moon by sending the signal $x \in  \mathcal{C}$ satisfying the power constraint $P_t$. This transmission is uniquely facilitated by a RIS composed of \(M\) elements, each endowed with the capability to adjust the signal's phase via shifts denoted by \(\phi_{m}\) for the \(m\)-th element. The RIS's function, captured in the phase shift matrix \(\bTheta_{ris} = \diag(\phi_{1}, \ldots, \phi_{M})\), \( \phi = e^{j\psi}\),   
    is pivotal in directing and enhancing the signal's power toward its lunar destination. This journey unfolds in two principal stages: the signal's initial travel to the RIS and its subsequent redirection to the Moon. 
    
    Overall, the signal model is given by \cite{cetin2023secure}:
    \[ y= \sqrt{P_r}\mathbf{h}^H \bTheta_{ris} \mathbf{g}x + n = \sqrt{P_r}\sum_{i=1}^M h_i^* \phi_i g_i x + n,\] 
    where $\mathbf{g} \in \mathcal{C}^{M \times1}$ is the channel from ground base station to the RIS, $\mathbf{h}^H \in \mathcal{C}^{1 \times M}$ is the channel from RIS to the destination, and $n$ is the additive white Gaussian noise with variance $N$. Moreover, $P_r$ is the abbreviation of $P_r(\mathbf{h}^H, \bTheta_{ris}, \mathbf{g})$, which represents the received signal power as the function of the channels, $\bTheta_{ris}$, and the path losses to be defined later. The corresponding SNR is then given by
\begin{equation}\label{eqn:snr_def}
\footnotesize
       \text{SNR} =  \frac{P_r(\mathbf{h}, \bTheta_{ris}, \mathbf{g})}{N}.
\end{equation}

\subsection{Path Loss Model}

    The path loss model involves the distances from Earth to the RIS \(d_{e,r}\) and from the RIS to the Moon \(d_{r,m}\). Assuming a free-space path loss model for simplicity, the path loss for each of the two links can be represented as follows \cite{jeong2022improved}:

i) From Earth to RIS:
$\footnotesize
PL_{e,r} = \left(\frac{4 \pi d_{e,r} }{\lambda}\right)^2,
$
 and ii) From RIS to Moon:
$\footnotesize
PL_{r,m} = \left(\frac{4 \pi d_{r,m} }{\lambda}\right)^2,
$
where \(\lambda = \frac{c}{f}\) with \(f\) being the frequency of the transmitted signal and \(c\) being the speed of light in a vacuum. Considering the RIS, the overall path loss for the signal from the ground base station to the Moon would be the product of these two path loss components, assuming they are independent.

Since all the links involved in such Cislunar communication systems are LoS whose signal attenuations are already captured by path losses, we may assume that $|\mathbf{h}|=|\mathbf{g}|=1$; therefore, only their phases matter. The calculation of the received power can then be simplified by computing the effective area \cite{bjornson2024introduction} as
 \begin{equation}
 \footnotesize
     \begin{split}
          A_{eff}(\mathbf{h}^H, \bTheta_{ris}, \mathbf{g})  &=  k \cdot A_{ris} \cdot \textrm{cos}^2(\bTheta_{opt}(\mathbf{h}, \mathbf{g}) - \bTheta_{ris}), \\
&= \sum_{i=1}^M k \cdot a_{ris,i} \cdot \textrm{cos}^2(\phi_{opt} - \phi_{ris,i}),
     \end{split}
 \end{equation}
where the directivity gain related to the area of the RIS is
 \(k \cdot A_{ris}\) with \(k\) being a constant that translates the area into directivity gain, and is based on fundamental principles of antenna theory. Meanwhile, the efficiency of phase adjustments 
\(\cos^2(\bTheta_{opt}(\mathbf{h}, \mathbf{g}) - \bTheta_{ris})\) highlights the RIS's ability to align its phases to focus the signal, where 
\(\bTheta_{ris}\) represents the deviation from an optimal phase alignment \(\bTheta_{opt}(\mathbf{h}, \mathbf{g})\). This cosine-squared function depicts the decline in efficiency as the phase alignment moves away from optimal.  
Overall, the received power $P_r$ can be derived as
\begin{equation}
\footnotesize
P_{\text{r}} =  \frac{P_t G_t G_r \lambda^2 A_{eff}(\mathbf{h}, \bTheta_{ris}, \mathbf{g})}{(4\pi )^3 d_{e,r}^2 d^{\prime 2 }_{r,m} } .
\end{equation}
Plugging this into \eqref{eqn:snr_def} shows that the received SNR is

\begin{equation}
       \text{SNR} =  
        \frac{P_t G_t G_r \lambda^2 A_{eff}(\mathbf{h}, \bTheta_{ris}, \mathbf{g})}{(4\pi )^3 d_{e,r}^2 d^{\prime 2 }_{r,m} N}.
\end{equation}

    This equation is fundamental in assessing the quality of the communication link, where a higher SNR indicates a clearer signal with less noise interference. In the next section, we formulate a constrained optimization problem that aims at maximizing the received SNR by optimizing $\bTheta_{ris}$.

    \section{Problem Formulation and Proposed Solution }\label{sec_meth}

 In this section, we introduce the problem as described in Section \ref{subsec_problem} and outline the solution in Section \ref{subsec_admmso}.

    \subsection{The Problem}\label{subsec_problem}

    The optimization problem for Cislunar communication networks aims to maximize the SNR by strategically configuring RIS phase \(\Theta_{ris}\). The focus is on maintaining reliable communication over long distances by achieving the required SNR to ensure signal quality under specific environmental conditions, such as free-space loss \(L_{fs}\) and noise \( P_{\text{noise}} \), thus ensuring efficient and sustainable operations for Cislunar destinations.

  The SNR equation underscores the critical importance of precise RIS phase rotation management to improve Cislunar communications. The subsequent formulation demonstrates how RIS technology can enhance the reach and reliability of these networks,
\begin{equation}\label{eq_main_rev} 
\begin{split}
&\max_{ \bTheta_{ris}} \quad SNR ( \bTheta_{ris})\\
& \text{subject to} \quad 
 \mathsf{C}_1: 0\leq \psi \leq 2\pi , \\
&\hspace{0.69in} \mathsf{C}_2: 0 \leq P_t \leq P_{t, \max},
\end{split}
\end{equation}
where the optimization constraints for enhancing wireless communication systems include constraint $\mathsf{C}_1$, which limits the minimum and maximum values for RIS phase rotation, and constraint $\mathsf{C}_2$, which governs the transmit power.

   \subsection{Problem Solution}\label{subsec_admmso}

The maximization of SNR in RIS systems involves addressing the phase rotation of the RIS.
 We assume the effective area of the RIS, \(A_{ris}\), is fixed at its maximum value, \(A_{max}\), which is fixed in the deployment phase and not the concern at this stage. 

The critical factor for SNR maximization is the phase rotation, controlled through the RIS phases. Recall that 
\[
A_{eff} = \sum_{i=1}^M k \cdot a_{max,i} \cdot \cos^2(\phi_{opt} - \phi_{ris,i}),
\]
where \(\phi_{opt}\) is the phase angle that maximizes the SNR and \(\phi_{ris,i}\) represents the phase set on each RIS element.

To achieve maximum SNR, the cosine squared term must be maximized. 
This is achieved when:
\[
\cos^2(\phi_{opt} - \phi_{ris,i}) = 1.
\]

This condition is satisfied when \(\phi_{opt} - \phi_{ris,i}\) is an integer multiple of \(2\pi\), leading to:
\[
\phi_{ris,i} = \phi_{opt} + 2\pi n, \quad n \in \mathbb{Z}.
\]

For practical purposes and simplicity, the most straightforward implementation sets \(n = 0\), leading to
\[
\phi_{ris,i} = \phi_{opt}.
\]

Therefore, the optimal \(\bTheta_{ris}\) matrix, which is a diagonal matrix of RIS phase angles, should be set to \(\bTheta_{opt}\) for each element. We note that in such a Cislunar mission, determining \(\bTheta_{opt}\) is a geometry problem based on the relative positions of the Earth transmitter, RIS on the GEO, and the Moon. This is because all the links involved, including that from the Earth transmitter to the RIS and that from RIS to the Moon, are LoS. We thus assume that there is a real-time adaptive control system to keep track of this phase; hence, $\phi_{opt}$ is known. 

To find the closed-form solution for the optimal angle \(\phi\), we assume that the positions of the ground station, the RIS-mounted GEO satellite, and the Moon vary in a three-dimensional space. We define their general positions and then derive the equations needed to calculate the optimal angle \(\phi\), which is the angle between the vector from the ground station to the GEO satellite and the vector from the GEO satellite to the Moon.
Define general positions  as:1)  Ground Station on Earth, coordinates \((x_e, y_e, z_e)\),
    For simplicity, we assume this is at the origin: \((0, 0, 0)\). 
   2) RIS-mounted GEO Satellite, coordinates \((x_{\text{geo}}, y_{\text{geo}}, z_{\text{geo}})\),
    GEO satellites are typically positioned above the equator at a fixed longitude. We assume the satellite is located at \((0, 0, 35786)\) km, and 
 3) Moon, coordinates \((x_m, y_m, z_m)\),
    We assume a general position in space, varying with time as the Moon orbits the Earth.


To calculate the distances, we use the Euclidean distance formula. From Ground Station to GEO Satellite \((d_{e,r})\):
\[
d_{e,r} = \sqrt{(x_{\text{geo}} - x_e)^2 + (y_{\text{geo}} - y_e)^2 + (z_{\text{geo}} - z_e)^2},
\]

From GEO Satellite to the Moon \((d_{r,m})\):
\[
d_{r,m} = \sqrt{(x_m - x_{\text{geo}})^2 + (y_m - y_{\text{geo}})^2 + (z_m - z_{\text{geo}})^2},
\]
We compute the angle \(\phi\) between the vector from the ground station to the GEO satellite and the vector from the GEO satellite to the Moon using the dot product formula.

Let \(\vec{v}_{e,r}\) be the vector from Earth to the GEO satellite:
\[
\vec{v}_{e,r} = (x_{\text{geo}} - x_e, y_{\text{geo}} - y_e, z_{\text{geo}} - z_e) = (x_{\text{geo}}, y_{\text{geo}}, z_{\text{geo}})
\]

Let \(\vec{v}_{r,m}\), be the vector from the GEO satellite to the Moon:
\[
\vec{v}_{r,m} = (x_m - x_{\text{geo}}, y_m - y_{\text{geo}}, z_m - z_{\text{geo}}),
\]

The dot product of \(\vec{v}_{e,r}\) and \(\vec{v}_{r,m}\), is:
\[
\vec{v}_{e,r} \cdot \vec{v}_{r,m} = x_{\text{geo}}(x_m - x_{\text{geo}}) + y_{\text{geo}}(y_m - y_{\text{geo}}) + z_{\text{geo}}(z_m - z_{\text{geo}}),
\]

The magnitudes of the vectors are:
\[
|\vec{v}_{e,r}| = \sqrt{x_{\text{geo}}^2 + y_{\text{geo}}^2 + z_{\text{geo}}^2},
\]
\[
|\vec{v}_{r,m}| = \sqrt{(x_m - x_{\text{geo}})^2 + (y_m - y_{\text{geo}})^2 + (z_m - z_{\text{geo}})^2},
\]

Therefore, the cosine of \(\phi\) is given by:
$
\cos(\phi) = \frac{\vec{v}_{e,r} \cdot \vec{v}_{r,m}}{|\vec{v}_{e,r}| \, |\vec{v}_{r,m}|},
$
solving for \(\phi\):
\[
\phi = \cos^{-1}\left(\frac{\vec{v}_{e,r} \cdot \vec{v}_{r,m}}{|\vec{v}_{e,r}| \, |\vec{v}_{r,m}|}\right).
\]

   The optimal reflection angle \(\phi\) is vital for directing RIS-reflected signals effectively toward the Moon, especially under misalignment with GEO satellites. Real-time RIS adaptation and phase tuning based on lunar motion and link dynamics are essential to maximize signal strength and SNR.

    \section{Simulation Results}\label{sec_sim}

    In this section, we evaluate the performance of our proposed model through extensive simulation results. Specifically, in Section~\ref{subsec_setting}, we provide the simulation setting, followed by the simulation results in Section~\ref{subsec_results}.

    \subsection{Simulation Setting}\label{subsec_setting}
The simulation scenario of the RIS-assisted Cislunar communication system, in which the RIS is mounted on a GEO satellite, is illustrated in Fig.~\ref{subfig:STK1a}, and the corresponding simulation parameters are summarized in Table~\ref{table:parameters}.

\begin{table}[!ht]
\scriptsize
\caption{Simulation Parameters}
\centering
\begin{tabular}{|p{1.0cm}|p{5.8cm}|p{0.8cm}|}
\hline
\textbf{Parameter} & \textbf{Value/Description} & \textbf{Ref} \\
\hline
\( P_{t } \) & Transmit power, 40 kW & \cite{9945839} \\
\( P_r \) & Calculated based on the equation provided & --- \\
\( G_t \) & Gain of the transmitting antenna, e.g., 30 dBi & \cite{donmez2023mitigation} \\
\( G_r \) & Gain of the receiving antenna, e.g., 20 dBi & \cite{donmez2023mitigation} \\
\( L_{ris} \) & Additional path loss due to RIS, e.g., 0.9 & \cite{9206044} \\
\( N \) & Noise level at the receiver, e.g., -100 dBm & \cite{jesick2020mars} \\
\( \lambda \) & Wavelength of the transmitted signal, e.g., 0.03 m & \cite{donmez2023mitigation} \\
\( \gamma_{th} \) & SNR threshold, e.g., 2 dB & \cite{jesick2020mars} \\
\( A_{max} \) & Maximum effective area of RIS, e.g., 100 m\(^2\) & \cite{9206044} \\
\( \psi \) & Between 0 and 2\(\pi\) & \cite{9206044} \\
\( k \) & Constant for directivity gain calculation, e.g., 0.1 & \cite{jesick2020mars} \\
\( \phi_{opt} \) & Various depending on defined factors & --- \\
\hline
\end{tabular}
\label{table:parameters}
\end{table}

    \subsection{Simulation Results}\label{subsec_results}

\begin{figure}[t]
	\includegraphics[width=0.8\linewidth]{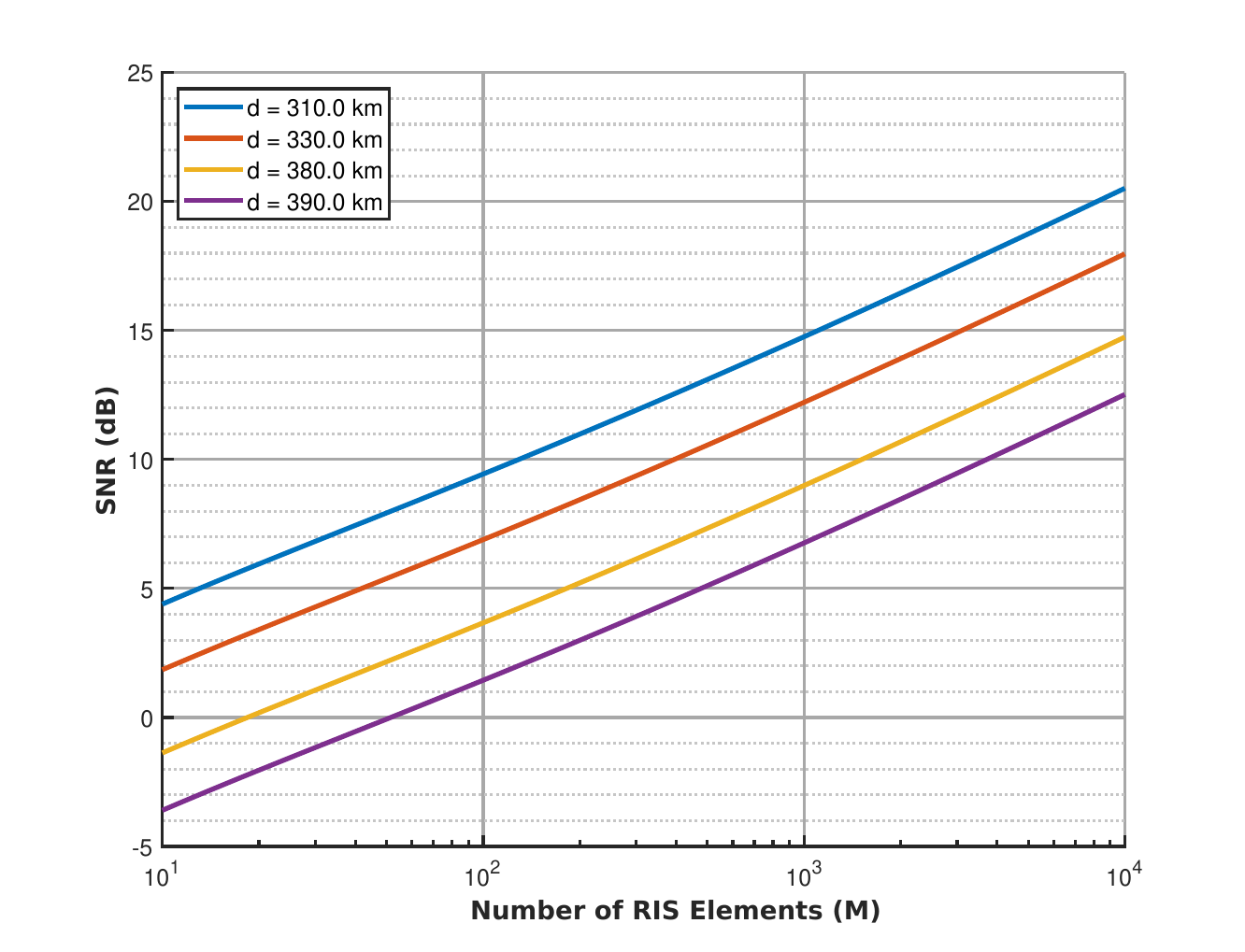}
    \caption{  SNR vs Number of RIS Elements }
    \label{fig:TVU_SIMF12}
    \end{figure}

Fig.~\ref{fig:TVU_SIMF12} illustrates the relationship between the SNR and the number of RIS elements in a Cislunar communication system. The results show that increasing the number of RIS elements leads to a higher SNR, which is expected due to the RIS-induced gain. This increase continues until a saturation point is reached. Simulations were conducted for various Earth–Moon distances $d$, with the highest SNR observed at the shortest distance, which aligns with the path loss effect shorter distances experience less signal attenuation. The RIS significantly enhances beamforming gain and signal reflection efficiency, thereby compensating for the severe path loss over the Earth–Moon link. However, the SNR improvement begins to plateau beyond a certain number of RIS elements due to practical limitations such as hardware constraints and the dominance of the noise floor. These results support the feasibility of deploying RIS on top of a GEO satellite to enhance Cislunar communication. Nevertheless, further investigation is needed to determine the optimal placement and configuration of the RIS in space.

    \begin{figure}[ht]
	\includegraphics[width=0.8\linewidth]{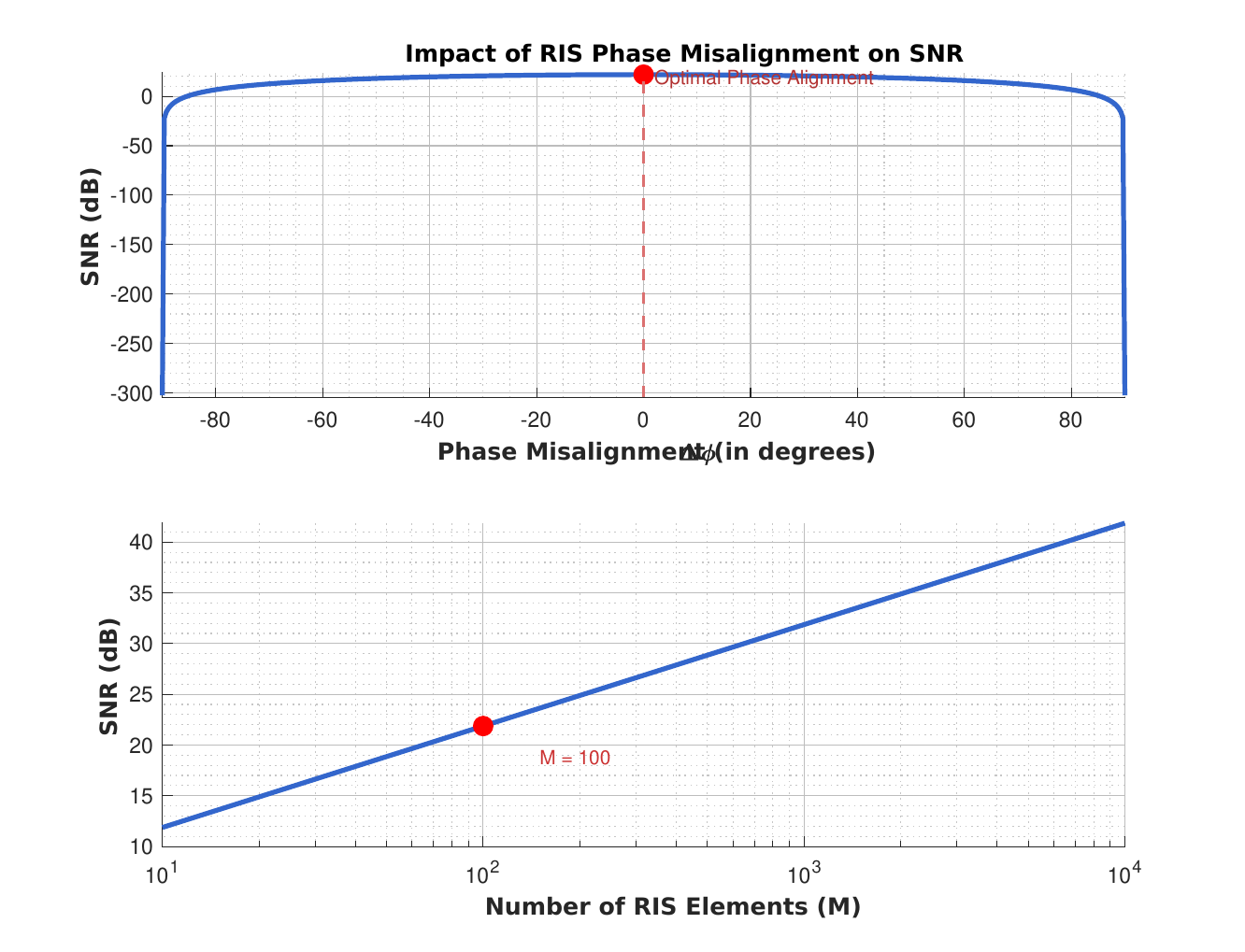}
    \caption{Impact of RIS Phase Misalignment}
    \label{fig:SIMS3}
    \end{figure}
The simulation results in Fig.~\ref{fig:SIMS3} demonstrate the dual impact of RIS phase misalignment and RIS size on the SNR, which are critical to the effective deployment of RIS in wireless communication systems. In the top subplot, the SNR reaches its maximum when RIS elements are perfectly aligned (i.e., $\Delta \phi = 0^\circ$), enabling constructive interference at the receiver. However, even small deviations in phase alignment lead to a significant drop in SNR, highlighting the system’s sensitivity to phase errors caused by hardware imperfections, environmental changes, or outdated channel state information. Once a certain misalignment threshold is crossed, destructive interference dominates, causing the SNR to fall below $-200$ dB, which renders the system practically unusable. This underscores the importance of accurate and real-time phase tuning. Meanwhile, the bottom subplot shows a monotonic increase in SNR with the number of RIS elements $M$, attributed to enhanced beamforming gain and improved wavefront control. Substantial gains are observed starting from $M = 100$, with continued improvement up to $M = 10^4$, indicating strong scalability. Nonetheless, such gains must be weighed against practical trade-offs, including increased power consumption, deployment cost, and the computational complexity of channel estimation. Therefore, both phase precision and scalable system design are essential for maximizing the benefits of RIS-assisted Cislunar communication networks.

    \section{Conclusion}\label{sec_con}
    
 This study introduced an innovative methodology to enhance Cislunar communications by utilizing RIS. It addressed challenges such as signal attenuation and the need for precise directivity and signal convergence, which are critical in Cislunar communication systems. The research focused on the optimization of the SNR for signals transmitted between the Earth and the Moon. Through the derivation of a closed-form solution, the optimal phase configuration for RIS was identified, leading to a substantial improvement in signal clarity. The findings revealed the significant benefits of adopting optimal RIS configurations. Potential future works include expanding to include diverse RIS-assisted communication scenarios such as multiple antenna systems, investigating various RIS technological architectures such as BD-RIS, and employing machine learning algorithms to refine RIS configurations further. In addition, future studies can explore the integration of RIS with lunar orbiters to enable dynamic relay systems. Investigating RIS-assisted full-duplex communication to enhance spectral efficiency in Cislunar networks also offers promise.

\section*{Acknowledgment}
This work has been partially funded by the National Science and Technology Council under the Grants NSTC 114-2221-E-A49 -185 -MY3, and NSTC 113-2634-F-A49-007, and NSTC 113-2218-E-A49-027 -, NSTC 113-2321-B-A49-019 -,Taiwan. 
This work was supported by the Higher Education Sprout Project of the National Yang Ming Chiao Tung University and Ministry of Education (MOE), Taiwan. Also supported in part by NSTC grants 114-2914-I-A49-021-A1 and 114-2811-E-A49-516-MY3.

    \bibliographystyle{IEEEtran}
    \bibliography{Ref.bib}
    \balance
    \end{document}